# Implications of Electronics Constraints for Solid-State Quantum Error Correction and Quantum Circuit Failure Probability

James E. Levy, Malcolm S. Carroll, Anand Ganti, Cynthia A. Phillips, Andrew J. Landahl, Thomas M. Gurrieri, Robert D. Carr, Harold L. Stalford, Erik Nielsen

*Abstract*— In this paper we present the impact of classical electronics constraints on a solid-state quantum dot logical qubit architecture. Constraints due to routing density, bandwidth allocation, signal timing, and thermally aware placement of classical supporting electronics significantly affect the quantum error correction circuit's error rate. We analyze one level of a quantum error correction circuit using nine data qubits in a Bacon-Shor code configured as a quantum memory. A hypothetical silicon double quantum dot quantum bit (qubit) is used as the fundamental element. A pessimistic estimate of the error probability of the quantum circuit is calculated using the total number of gates and idle time using a provably optimal schedule for the circuit operations obtained with an integer program methodology. The micro-architecture analysis provides insight about the different ways the electronics impact the circuit performance (e.g., extra idle time in the schedule), which can significantly limit the ultimate performance of any quantum circuit and therefore is a critical foundation for any future larger scale architecture analysis.

This work was supported by the Laboratory Directed Research and Development program at Sandia National Laboratories. Sandia is a multiprogram laboratory operated by Sandia Corporation, a Lockheed Martin Company, for the United States Department of Energy's National Nuclear Security Administration under Contract DE-AC04-94AL85000.

J. E. Levy is with the Sandia National Laboratories, Albuquerque, New Mexico 87185 USA. (e-mail: jelevy@sandia.gov ).
M. S. Carroll is with the Sandia National Laboratories, Albuquerque, New Mexico 87185 USA. (e-mail: mscarro@sandia.gov ).
A. Ganti is with the Sandia National Laboratories, Albuquerque, New Mexico 87185 USA. (e-mail: aganti@sandia.gov ).
C. A. Phillips is with the Sandia National Laboratories, Albuquerque, New Mexico 87185 USA. (e-mail: caphill@sandia.gov ).
A. J. Landahl is with the Sandia National Laboratories, Albuquerque, New Mexico 87185 USA. (e-mail: ajland@sandia.gov ).
T. M. Gurrieri is with the Sandia National Laboratories, Albuquerque, New Mexico 87185 USA. (e-mail: tmgurri@sandia.gov ).
R. D. Carr is with the Sandia National Laboratories, Albuquerque, New Mexico 87185 USA. (e-mail: rdcarr@sandia.gov ).
H. L. Stalford is with the Sandia National Laboratories, Albuquerque, New Mexico 87185 and with the School of Aerospace and Mechanical Engineering, University of Oklahoma, Norman, OK 73019 USA (e-mail: hlstalf@sandia.gov and stalford@ou.edu ).
E. Nielsen is with the Sandia National Laboratories, Albuquerque, New Mexico 87185 USA. (e-mail: enielse@sandia.gov ).

## I. Introduction

Quantum computation has attracted significant attention for its potential to address problems that are intractable with classical computing approaches. A building block of the hypothetical quantum computer is the quantum bit (qubit), which stores information in the probability amplitudes of a two level quantum system. The qubit information is, however, not perfectly isolated from its environment leading to eventual errors on the qubit due, for example, to decoherence of the qubit information. A breakthrough in the field of quantum computing was the realization that errors in the qubit information could be detected and corrected through quantum error correction (QEC) codes [1-3].

Solid-state approaches to quantum computing have attracted attention for numerous reasons including recent successes in experimentally demonstrating a qubit [4-6], the promise of potentially long decoherence times in silicon [7-9], the potential for fast and universal operations relative to the decoherence time [10] and the existing solid-state infrastructure for circuits and digital logic. Analyses of solid-state architectures often attempt to identify the requirements for qubit performance (i.e., the probability of a single gate error) that is sustainable by the QEC choice. QEC codes are designed to correct a certain number of errors. Fewer errors must occur during all the operations at the coding and checking to show benefit.

A fundamental problem in quantum information science and engineering is to develop a realistic description of the classical interface to the qubit and quantify its impact on the performance, in this case a QEC code. To date, some quantum computing architecture analyses have attempted to include the impact of the classical interface for the solid-state system [11-13] but the treatment of the interface is typically peripheral and commonly the assumptions made about the classical interface obscure important guidance about this critical element of quantum circuitry. In this paper, we will highlight several elements of the classical interface to small quantum circuits that define system parameters of the micro-architecture, such as the minimum quantum clock period and scheduling penalties (e.g., reduced parallelism) that have implications at the micro-architecture level and for solid-state quantum circuits in general.

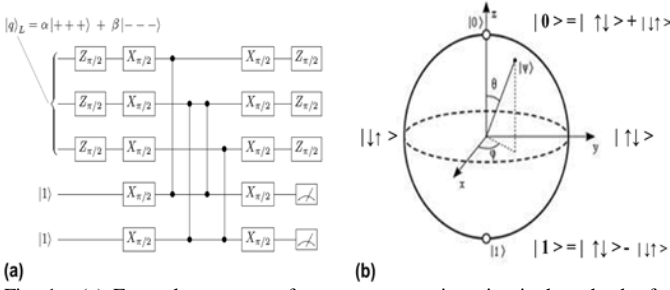

Fig. 1. (a) Example segment of an error correction circuit that checks for errors after the qubit information has been encoded into 3 qubits. (b) Bloch sphere and definition of |0> and |1> states in the two spin basis of a 2 electron double quantum dot.

A quantum circuit, like the one shown in Fig. 1 (a), is an example of an abstract representation of one part of a complete QEC code. A qubit's state, the probability amplitudes of the |0> and |1>, can be described with the angular direction of a unit vector on a spherical surface called the Bloch sphere, Fig. 1 (b). Single qubit operations can be described as rotations on the Bloch sphere ($Z_{\pi/2}$, $Z_\pi$, $X_{\pi/2}$). More complex operations that couple two qubits together (CPHASE) are indicated as the vertical lines between the qubits in Fig. 1 (a). The horizontal lines represent the timeline of a single qubit with time advancing from left to right. To execute the error correction circuit a sequence of gate operations must be scheduled at the abstract level of qubit operations, indicated by the error check circuit in Fig. 1(a) [14].

Each of these qubit rotations represents at least one and sometimes several high speed, high accuracy and high precision complex time varying voltage or current pulses sent to specific conducting lines of the physical qubit, Fig. 3(b). A more complete explanation of QEC and QEC circuits go beyond the scope of this paper and are described in significant detail elsewhere [14].

The absence of a well established long-range transport mechanism in solid-state qubits leads to a lay-out that relies on local interaction between the qubits. To accommodate this limitation, a local error correction code was chosen called the Bacon-Shor code [15]. A single qubit's information is encoded in nine data qubits and additional ancilla qubits are added to allow measurements to take place without directly interfering with the information in the data qubit as well as providing fault tolerance against propagating errors in the quantum circuit [1]. This leads to a lay-out of 21 total qubits, Fig. 3 (a). This code provides protection against one error (i.e., a distance 3 code).

Disregarding any constraints such as locality and transport the minimum number of gate operations required to execute an algorithm amounts to counting the operations in a circuit such as Fig. 1 (a). The introduction of electronics results in constraints which introduce scheduling conflicts. Some qubits, for example, must wait (i.e., incur idle ticks), while shared conducting routes are utilized by other qubits to perform operations. An example of a provably optimal schedule [16],[17], that shows each of the qubits, their operations and their required idle ticks is shown in Fig. 2. The electronics constraints described in the body of this paper were applied to this schedule. Dynamic decoupling pulses, often inserted to reduce decoherence during idle, are not included in this schedule.

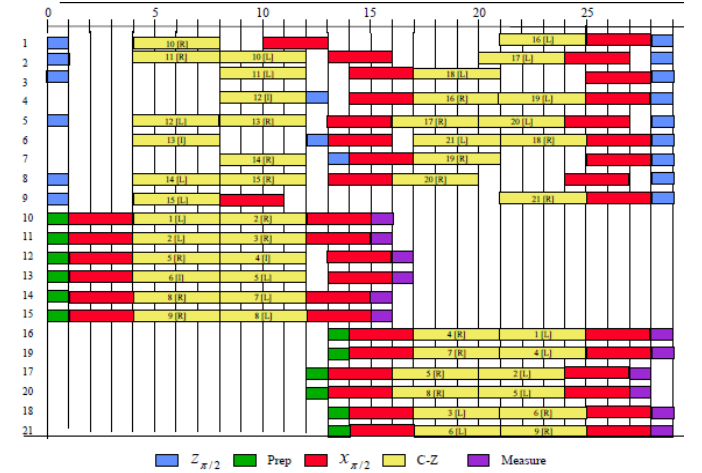

Fig. 2. A schedule for the proven optimal error correction code. With the total number of idles, q, being optimized.

A pessimistic probabilistic bound on the error rate of a quantum circuit provides some intuition about which factors affect the error correction performance. A gate error probability can be estimated below which the circuit provides benefit. If an error correction circuit has a combination of N qubit operations (i.e, gates) that err with probability p and the qubits must idle a sum of M clock periods with a probability of error q, then an error probability for a distance 3 error correction circuit, $p_{circuit}$, can be estimated as:

$$p_{circuit} \leq \frac{M(M-1)}{2}q^2 + (Np)(Mq) + \frac{N(N-1)}{2}p^2 \quad (1)$$

The error correction circuit provides some benefit when $p_{circuit}$ is less than p. A separation of idle error rates, q, and gate error rates, p, helps account for the impact of classical electronics. The electronics introduces possible limits on p because of limits in the accuracy or precision of the classical electronics available as well as introducing penalties in the schedule due to parallelism limits (i.e., how many qubits can be manipulated simultaneously). Schedule penalties are observed as an increase in the number of idle blocks for each individual qubit, M. The probability of an error during the smallest interval of time, q, might also be defined by limits in the way the circuit is clocked, although we note that this can also be counted as an increase in M idle steps. The error q, for example, might be estimated as the minimum clock period by $T_2$.

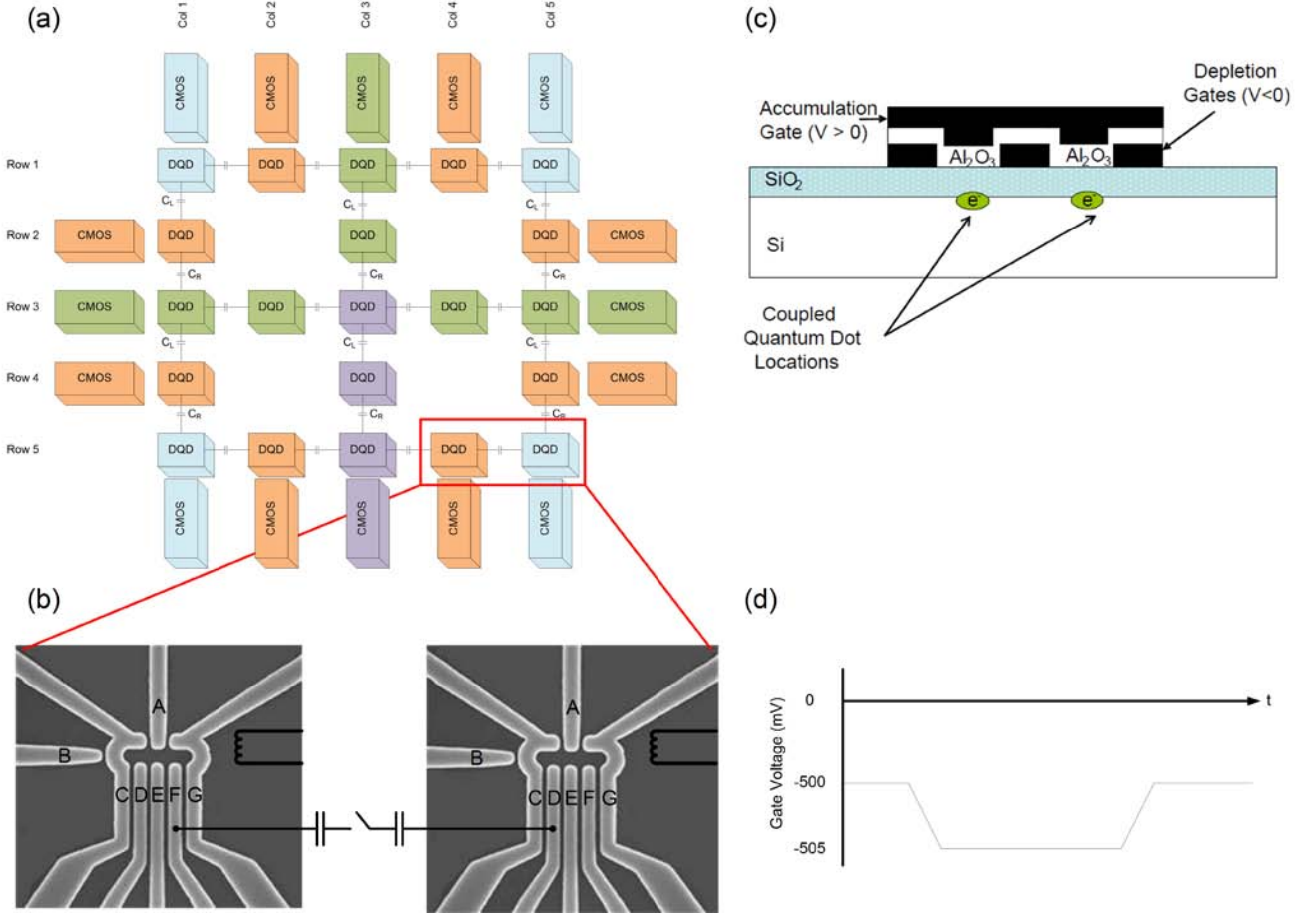

Fig. 3. (a) lay-out of qubit positions and surrounding classical electronics to route signals to the qubits at 100 mK. (b) scanning electron microscope image of silicon double quantum dot structure emulating GaAs qubit structure. Conceptual capacitance coupling and local inductor for the X-gate are also indicated. (c) cross section of silicon quantum dot device and (d) conceptual example of pulse sequence sent to gates to do a Z rotation.

The following sections of the paper discuss the role the electronics architecture plays in determining the circuit failure rate. The general ways the electronics impacts the circuit error rate are in three forms: 1) increased number of idle steps in the schedule; 2) definition of the minimum idle time and therefore the minimum idle error, q; and 3) increased errors in qubit operations/gates. Section II outlines a hypothetical silicon qubit and the electronics staging within a dilution refrigerator for a nine data and twenty one total qubit Bacon-Shor code, BS9(21). In section III the impact of the electronics constraints on the bandwidth and parallelism (e.g., idle time) are discussed. In section IV we examine the voltage and timing accuracy required to perform quantum gates on a silicon double quantum dot and compare them against classical noise sources due to charge-injection and cross-talk. Discussion and summary of the implications of the electronics I/O for quantum circuitry is found in section V and VI.

## II. ELECTRONIC COMPONENT REQUIREMENTS FOR SILICON DOUBLE QUANTUM DOTS & ELECTRONICS STAGING

Silicon spin qubits are predicted to have long decoherence times because of weaker spin-orbit interactions and the ability to engineer isotope enriched silicon crystals that have a net nuclear spin zero [7]. Long spin decoherence times in donor ensembles have been measured using electron spin resonance [7] and significant progress towards few electron silicon quantum dot qubits has also recently been made [18-22]. Furthermore, the ability to integrate silicon CMOS electronics for input and output signals is a critical long term advantage of silicon based qubits providing an established technology around which an analysis of the electronics impact on quantum computing micro-architecture can be performed.

The physical qubit considered here is a silicon double quantum dot (Si-DQD). The Si-DQD qubit, Fig. 3 (b), uses a global top gate to accumulate electrons combined with depletion gates (labeled A-G), negative voltages, to confine and isolate single electrons within the quantum dot regions shown in Fig. 3 (c) [23], [18]. Rotations around the Z and X axes in the Bloch sphere can be achieved by changing the exchange energy and the gradient of the magnetic field along the axis of the DQD, respectively. A constriction near the double quantum dot region provides a single charge sensitive conducting channel whose conductance is used as a read-out of the qubit state. This native gate set is similar to a previous proposal by Taylor et al. for the GaAs DQD system [11] and many of the analogous single qubit operations have been demonstrated in GaAs at electron temperatures of approximately 100 mK achieved in a dilution refrigerator [10].

To date no Si-DQD qubit has been demonstrated although a number of precursor experiments have been reported including controlled single spin isolation and spin read-out [18-20]. The rapid progress in silicon single spin devices motivates this analysis. For the silicon system, we propose that an external gradient field will be supplied by a local inductor, Fig 2 (b), placed above the global accumulation gate. Coulomb interaction between polar positioning of the electron charge within the DQDs is proposed for the CPHASE [11]. It is assumed that a series of MOSFET switches can modulate the coupling capacitance sufficiently to make the CPHASE coupling negligible when two qubit operations are not desired. We note that both the CPHASE gate and the switch are speculative mechanisms. However, this assumption is not critical for the overall trends highlighted in this paper. The proposed Si-DQD has 15 conducting gates and ohmic contacts per qubit. Each of these conducting lines are connected to a classical electronics input or output that provide 1) fixed voltage biases; 2) high speed lines for voltage or current signals as well as read-out that define qubit rotations; and 3) B-fields from local inductors. In addition to the 15 gates and contacts described above 24 signals are required to control the MOSFET switches that modulate the CPHASE coupling capacitance. For a BS9(21) qubit quantum error correction circuit a total of 339 lines, 15 lines per qubit in addition to 24 control lines for the switches, must be energized by a combination of pulse generators, read-out circuitry and constant bias sources.

This is more lines than are typically available in standard commercial dilution refrigerators, which often have less than 100 lines available [24]. The total number of available lines in a dilution refrigerator is limited by space as well as concerns about maintaining a high level of thermal and noise insulation between warmer stages and the 100 mK stage since the 100 mK stage has a limited cooling power. Two options available to accommodate the limited number of routing lines between temperature stages are: (a) to integrate all electronics at the qubit temperature stage or (b) stage electronics at higher temperature stages and rely on multiplexing of signals. A simple estimate of power required just for pulse generation, based on work presented in [25] would require at least ~1.2 mW and likely greater than 100mW to meet the speeds and noise performance needed for quantum computation. The maximum cooling power of the lowest temperature stage of many commercial dilution refrigerators is less than 1mW, (e.g., 400 μW [24]), suggesting that reliance on multiplexing and staging of some or most electronics at higher temperature stages, where cooling power is greater, will be necessary. A suggested design that addresses the thermal cooling (400 μW) and a signal line limit of ~64 which is available commercially[24], is partitioned as: (a) a 300K stage that holds the master CPU for the system control and the pulse generators to drive the physical qubits, (b) a 4K stage that houses the digital readout circuitry in order to reduce the parasitic load between the qubits and read-out circuitry, and (c) a 100mK stage that holds the 21 physical qubits and supporting signal routing electronics (multiplexers (MUX's), de-multiplexers (DEMUX's), and memory to hold the state of the MUX's/DEMUX's.

The staging of the pulse generation electronics at room temperature in contrast to an intermediate temperature stage is motivated by two factors: (a) pulse generation must be moved off of the 100 mK stage due to heating concerns and (b) high performance pulse generation requiring sub-nanosecond resolution and wide dynamic range is best met by not limiting power and space requirements (i.e., room temperature staging). Fast signal injection can be achieved from room temperature using co-axial cables that run between 100 mK and room temperature with minimal performance penalty and has been demonstrated experimentally for all necessary single qubit rotations in GaAs [4],[10], although the X rotations relied on a built-in nuclear field that is not available in silicon. The pulse generators are therefore configured to provide multiple pulse sequences for any gate in our basis and the number of pulse generators is dictated by the number of CMOS control blocks in the system. There are 16 control blocks for the 21 qubits as arranged in Fig. 3 (a). A single co-axial cable down to 100 mK is dedicated to each pulse generator.

A program to execute the QEC circuit and the specific protocols for pulses to actuate the qubit rotations must be stored in a memory and transferred, when needed, to the pulse generators. For a quantum circuit there are a limited number of single and two qubit rotations making a limited set of pulse sequences that must be programmed for the qubits. The total number of different gate operations for the BS9(21) are listed in Table I. They require 7 different protocols (CPHASE requires 3 separate protocols) a limited set needed for the memory, for which only a few of the conducting gates on the physical qubit are actually changed during the operation. The protocols and program are stored in a memory that we envisage also being located at room temperature.

TABLE I
SUMMARY OF GATE COUNT FOR A SCHEDULE OF A HALF-ROUND OF QEC WITH AND WITHOUT ELECTRONICS CONSTRAINTS

| Gates | No Electronics Constraints | Electronics Constraints |
|---|---|---|
| Prep | 12 | 12 |
| X$\pi$/2 | 42 | 42 |
| Z$\pi$/2 | 18 | 18 |
| CPHASE | 24 | 24 |
| Msr | 12 | 12 |
| **Idle (Ticks)** | ***48*** | ***95*** |

The placement of the read-out electronics is complicated by potential penalties in speed of response, noise performance, or both when staged at higher temperatures. In order to reduce the delay associated with long wires within the cryostat, readout circuitry can be placed at a 4K stage, as close as possible while not violating the power budget associated with the cryostat. The closer proximity and colder device temperature in principle should provide both noise and bandwidth benefits.

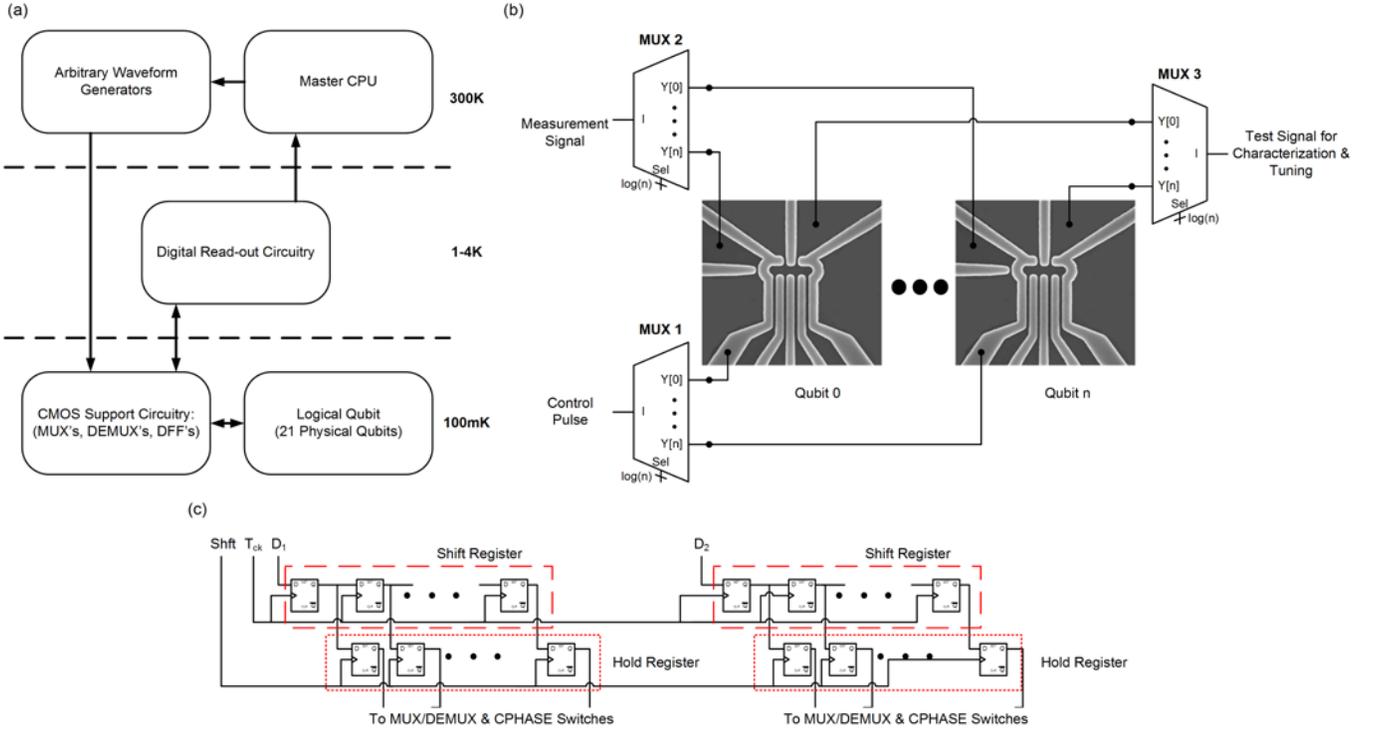

Fig. 4. (a) Electronics staging of quantum circuit design and (b) Configuration of memory elements located at 100mK to serialize control information. (c) Schematic diagram of MUX and DEMUX used by CMOS blocks controlling multiple qubits in the BS9(21) architecture.

## III. BANDWIDTH, ROUTING, & LAYOUT

### A. Definition of signal lines between stages

A multiplexed configuration is used because there are not enough lines to directly address every one of the 21 qubit's control lines. The multiplexing leads to a loss of parallelism for qubit operation and an increase in idle time. The increase is manifested in an increased number of idle ticks, M in equation (1), because a fraction of qubits must wait idle while the others are gated. The configuration of high speed and DC signal lines, therefore, is important in determining the parallelism penalty. We find that the following distribution of lines meets the functional needs of the 21 qubit system, which includes the following configuration of routes between room temperature and the 100 mK level:

- 16 signal lines dedicated to each of the arbitrary waveform generators
- 16 analog measurement lines from each CMOS block
- 22 lines to drive current to the local inductors including a shared ground
- 10 shared bias lines to all 21 qubits including a shared ground
- 1 line used for qubit characterization and tuning
- 4 signal lines used to serialize all MUX, DEMUX, and CPHASE switch control data.

The 16 lines allocated for waveform generators assumes that a single generator is dedicated to a CMOS block, some of which serve multiple qubits. This reduces lines down from room temperature but introduces a scheduling constraint that those shared qubits may either select that particular gate operation or must otherwise be idle during that time. Analog measurements sample the current from the charge sensor of the 100 mK qubit, which includes selecting the charge sensor through the corresponding CMOS block and processing the small currents at the 4K stage. One measurement output line, 16 total lines, service each of the CMOS blocks, which has the consequence that only one qubit can be measured in a single block at a particular time. A local inductor is placed near every qubit for the X-rotation and a dedicated superconducting line (e.g., niobium with a relatively high critical field) is needed for each of the 21 qubits. Some DC voltages can be shared across all the qubits if the qubits are identical, which is assumed for this analysis although it is likely that additional tuning schemes (e.g., customized pulses) will be necessary in a more realistic estimate of the system. Finally, the four signals used to send serialized control information are a clock signal ($T_{clk}$), shift signal (Shft), and 2 data signals ($D_1$, $D_2$). Two data signals are required to send information serially as a result of the 45 bits of control information that are needed (24 bits for CPHASE control, 21 bits for MUX/DEMUX control) and our choice of clock frequencies discussed in the next section. The arrangement of the memory elements (D Flip-Flops) relative to these signals is shown in Fig. 4 (c). In order to avoid control signals from toggling unintentionally while reading in a new control word both a shift register as well as a hold register are required.

## B. Definition of classical and quantum processor clock

Time in the system is defined by two time periods, the classical clock period, $T_{clk}$, and the quantum processor clock period, $T_{Qclk}$. The quantum clock period is the time of the shortest quantum gate operation and is the time segment in which the qubit operations are scheduled for the quantum circuit, Fig. 2. Each quantum clock period is broken into smaller time segments, which are defined by the speed of the classical processor (i.e., the classical clock period). The timing of the classical signal pulses, read-out and digital information between stages are all defined as increments of this classical clock period. Many sources of error (e.g., decoherence) in the qubit grow increasingly likely with time (i.e., q in equation (1) can be dependent on $T_{Qclk}$) making it important to minimize $T_{Qclk}$ (e.g., $q \sim T_{Qclk}/T_2$).

One potential limit on $T_{Qclk}$ can arise due to limited digital bandwidth for passing information between room temperature and the quantum processor stage for the selection of specific qubits and gate operations on those qubits. The total bandwidth that is available between the 100 mK stage and the other stages is dependent on: 1) the size and number of the MUXes or DEMUXes, which is dictated by the number of qubit lines and qubits controlled by a particular MUX/DEMUX; 2) the number of CPHASE switches; 3) the classical clock speed; and 4) the number of signal lines used to send serial control information. The size of the binary word that is necessary is a summation of all the MUX/DEMUX control lines as well as the signal used to control the CPHASE switches. The number of control bits for a particular MUX(DEMUX) is dependent on the size of the MUX(DEMUX). A pipelined data path is required in which an N bit wide binary word, is sent to program the MUX/DEMUX and CPHASE control bits for a time segment k. Furthermore, the N bits must be sent during the quantum gate being performed at time k-1 in order to setup the MUX/DEMUX serially in time for the next quantum clock period. This pipelining is achieved through the configuration shown in Figure 4 (b). Digital select line bandwidth becomes important in determining $T_{Qclk}$ because $T_{Qclk}$ can not be faster than the time it takes to transfer the digital data. $T_{Qclk}$ is, therefore, generally a function of the size of the system (i.e., number of qubits), the complexity of the qubit (i.e., number of signal lines to control the qubit), as well as the complexity of the algorithm (i.e., the number of different gate operation protocols necessary). Fig. 5 shows the required serial data signal lines (represented at $D_1$ & $D_2$ in Fig. 4 (b)) as a function of the $T_{Qclk}$ to $T_{clk}$ ratio, using a 45 bit set-up. The fastest classical CMOS clocks ($T_{clk}$) run at ~3 GHz indicating that a $T_{Qclk}$ of 15 ns is achievable with 1 serial data line in a 21 DQD qubit system. Faster $T_{Qclk}$ times can be arranged with more serial data lines, however, in our analysis we relax $T_{clk}$ to 1ns for reasons related to the electronics discussed in the next session. This results is a $T_{Qclk}$ of 45 ns for the BS9(21) with 1 serial data line or 23 ns with 2 serial data lines. We therefore use 2 data serial lines as shown in Fig. 4 (c) and choose $T_{Qclk}$ to be 30 ns providing a bit of pessimism in our analysis.

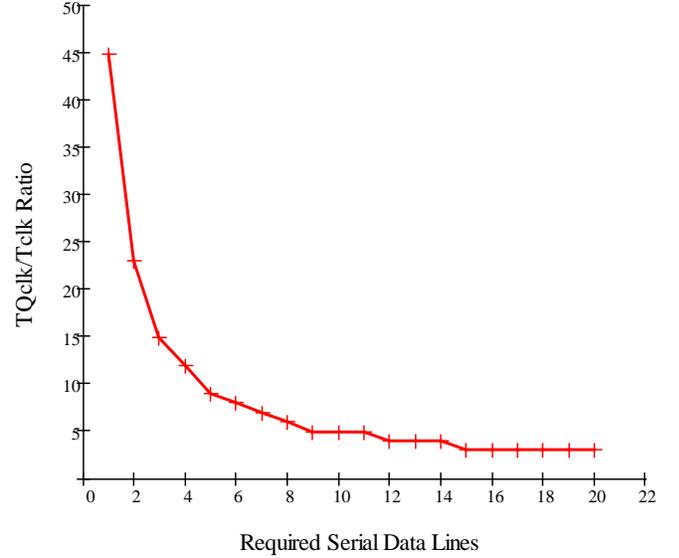

Fig. 5. Required serial data lines for BS9(21) architecture based on Quantum Clock to Classical Clock design target.

## C. Routing constraints on layout of qubits at 100 mK stage

Parallelism is impacted by layout of the qubits due to restrictions in parallel gating which are applied to reduce the effect of cross-talk between qubits. Standard transistor technologies can manufacture a limited number of metal layers producing a limited volume of space through which metal routes can run [12]. Each metal route has a minimum feature size designated by the technology, therefore, the metal can not be scaled to arbitrarily small sizes to compensate for space. A calculation can be done based on known manufacturing limits to establish the maximum number of routes, that is qubits, that can be reached along a length with a fixed width. The results for an optimistic scenario that assumes no additional space compensation or use of shielding for cross talk, Table II, show that even for advanced process nodes such as 45nm the number of accessible qubits is limited by the number of lines needed for each qubit and the width of the qubit footprint itself. It was assumed that no lines could be shared in this calculation, however, in some scenarios (such as high qubit matching) it may be possible to share the same DC bias voltages across multiple qubits which would increase the number of qubits that could be reached. The BS9(21) architectures assume high qubit matching and shares 9 DC bias voltages, which we emphasize is probably an optimistic assumption.

Table II
SUMMARY OF ROUTING LINES AND CONTROLLABLE QUBITS IN A 1D GEOMETRY WITH A FIXED WIDTH OF 1.5 PHYSICAL QUBITS (1.25μM)

|  | Process Technology Node | | | | |
| --- | --- | --- | --- | --- | --- |
|  | 350nm | 130nm | 90nm | 65nm | 45nm |
| Routing Channels | 4 | 19 | 27 | 40 | 62 |
| Controllable Qubits[a] | 0 | 1 | 2 | 3 | 5 |

[a]Qubit control lines are not shared in this calculation.

The proposed two qubit mechanism uses capacitive coupling to mediate Coulomb repulsion between the two double quantum dots [11]. Increasing the length of the line used for the capacitive coupling is beneficial from a routing space perspective, however, the increased length of the coupling bar also makes it more susceptible to noise due to cross-talk from nearby signals (i.e., larger capacitance to other neighboring signals). In order to reduce cross-talk on this coupling bar and to keep gate times short: 1) a minimal length is considered, and 2) a constraint is imposed on the BS9(21) architecture that signals may only be active over a qubit if that qubit is in a steady "park" location (a position within the charge stability diagram that is less susceptible to noise) or performing the same gate operation. This reduces the introduction of errors due to cross-talk but results in additional loss of parallelism among those qubits sharing a classical controller [17].

## IV. Voltage & Timing Accuracy

The non-idealities of the classical electronics control can be a source of dominant error for physical qubit operations especially when pressed to the limits of their accuracy and precision for very fast pulsing. The exchange interaction is recognized as one of the more vulnerable operations to noise and decoherence due to the surrounding electrostatic environment and controlling electronics. In particular, classical electronics accuracy is limited in both voltage (e.g., noise and drift) and time (e.g., jitter) and will result in errant rotation with a dependence of order $\partial \phi = \partial J(V)\left(\dfrac{t}{\hbar}\right) + \partial t\left(\dfrac{J(V)}{\hbar}\right)$. Error probabilities due to voltage fluctuations in an exchange gate are very sensitive to the dependence of the exchange energy on voltage, $J(V)$, and have been estimated for several cases [26-28]. The error rate dependence on the accuracy of the control electronics can also be identified using this previous formalism, which in turn suggests limits on the minimum time of the exchange gate in this proposed BS9(21) quantum circuit.

### A. Timing Accuracy

The sensitivity of the exchange gate to timing inaccuracies like jitter can be examined by calculating the exchange energy dependence on applied voltage. An exchange energy model for a Si-DQD has been developed for this BS9(21) system Fig. 6 [29]. We assume for the timing analysis that the voltage pulse applied is a perfect square pulse. Gate error rates are targeted initially to be less than $10^{-4}$, which leads to a percentage error in targeted time, $\dfrac{\partial t}{T_{gate}}$, of no greater than $10^{-2}$ [12]. Commercially available pulse generators can provide signals with total peak-to-peak jitter on the order of 10ps. In addition to uncertainty in pulse time, additional jitter can be introduced in the rising and falling edge of the classical control clock, which is typically on the order of 1% of the clock period (e.g., 10 ps). A starting estimate of $T_{gate}$ constrained to $10^{-4}$ error would therefore be targeted at $T_{gate}$ no shorter than ~ 1 ns and reports of very fast pulsing of DQD qubits, for example $T_{gate}$ ~ 120 ps [4], would likely require an extremely fast classical clock time with very low jitter to achieve the necessary gate accuracy. Compensated pulse techniques [30-32] (e.g., BB1) and optimal control [33-35] may also assist in improving these constraints.

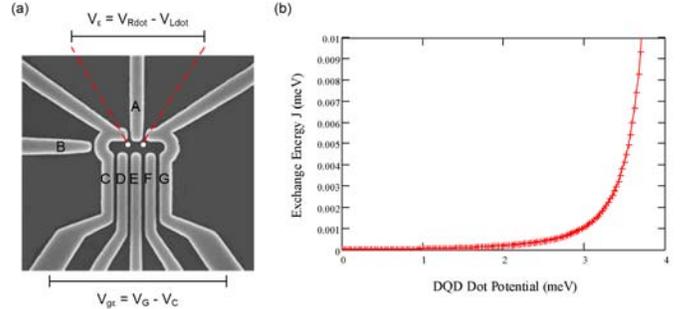

Fig. 6. (a) Si-DQD defining detuning voltage (b) Exchange Curve with

### B. Voltage accuracy and definition of $T_{Qclk}$

The required gate voltage accuracy for a $Z_\pi$ gate depends on the required error rate. The exchange gate rotation rate has a quasi-exponential dependence on applied voltage to the qubit, which has been measured in GaAs and calculated for the Si qubit for this design Fig. 6 [29]. To translate externally applied biases of hypothetical electronics sources to the theoretical quantum dot island potentials calculated in the exchange model, a small signal capacitance SPICE [36] model has been developed [37]. Using the capacitance model the effect of the voltage error at the dots ($V_\varepsilon$) was determined. The voltage $V_\varepsilon$ was then converted to an exchange energy J using the theoretical model of the Si-DQD [29]. Using the new exchange energy value, the gate rotation was calculated and compared to the ideal rotation ($Z_\pi$ for our analysis). The results are tabulated in Table III, where we see that because the exchange energy is exponential the magnitude of the error is dependent on both the size of the voltage error as well as where on the curve we are operating. Many quantum circuits require gate error rates less than a threshold (e.g., $10^{-4}$) in order to show benefit. A $10^{-4}$ error probability can be approximated as an error in rotation of approximately $10^{-2}$ [12]. A minimum gate time is therefore defined by the limits in the voltage accuracy of the electronics combined with the Z gate exchange energy, and sensitivity to voltage inaccuracy. $T_{Qclk}$ for this design is set to 30 ns to assure that voltage noise and timing jitter do not introduce worse than a $10^{-4}$ error probability, under the assumption that the supporting electronics and pulse generators can be designed with <$100\mu$V noise. This choice must be revisited if it is found that the probability of gate error must be less in order for the circuit to work. We also note that errors from other physical mechanisms such as random charge fluctuations within the qubit are not considered in this estimated error and that the error estimate is highly dependent on the exchange energy model which will fluctuate with magnetic field and bias conditions of the DQD.

**Table III**
COMPARISON OF ROTATION ERROR OF A $Z_\Pi$ GATE DUE TO ELECTRONICS NOISE FOR VARIOUS GATE TIMES

| J Target (µeV) | Gate Error (µV) | J Error (eV) | Z Error ($\pi$ radians) | Gate Time (ns) |
|---|---|---|---|---|
| .069 | 1 | 2.379E-11 | 1.0845E-03 | 30 |
|  | 10 | 2.383E-10 | 1.086E-02 |  |
|  | 100 | 2.418E-09 | 1.1019E-01 |  |
|  | 1000 | 2.735E-08 | 1.2464E+00 |  |
| .5 | 1 | 1.873E-10 | 1.1771E-03 | 4.13 |
|  | 10 | 1.878E-09 | 1.1799E-02 |  |
|  | 100 | 1.923E-08 | 1.2082E-01 |  |
|  | 1000 | 2.469E-07 | 1.5515E+00 |  |
| 1 | 1 | 4.757E-10 | 1.4945E-03 | 2.06 |
|  | 10 | 4.771E-09 | 1.4988E-02 |  |
|  | 100 | 4.910E-08 | 1.5427E-01 |  |
|  | 1000 | 6.756E-07 | 2.1225E+00 |  |
| 2 | 1 | 1.186E-09 | 1.8637E-03 | 1.03 |
|  | 10 | 1.191E-08 | 1.8701E-02 |  |
|  | 100 | 1.234E-07 | 1.938E-01 |  |
|  | 1000 | 1.879E-06 | 2.9518E+00 |  |

## V. DISCUSSION

The constraints imposed on electronics considered in this paper highlight both specific challenges for this quantum memory micro-architecture as well as general ways the electronics define the circuit performance. To calculate a quantitative estimate of the impact of the electronics, the circuit error rate is calculated using equation (1), both for a non-local schedule and for an optimal schedule for the quantum error correction code using the local architecture described in the previous sections. The error probability of the two circuits is calculated as a function of the gate error rate assuming a fixed idle error rate, q, in Fig. 7. The idle error rates of $10^{-4}$ (triangles), $10^{-5}$ (circles), and $10^{-6}$ (squares) are the same order of magnitude as would be expected for 30 ns idle steps, defined above by the classical electronics design, and literature reports of measured T2 decoherence times of electron spins in the bulk, 60 ms [7], and near an oxide surface, ~0.3 ms [9]. These are optimistic since they are produced using some form of dynamic decoupling, which is not included in this schedule. It is important to note that the 30 ns idle time is chosen to assure that the Z gate error probability is of order no greater than $10^{-4}$. For reference, the number of gates for both local and non-local optimal schedules [17] are shown in Table I.

The horizontal lines show the intersection of the bare idle error rate and the quantum error corrected memory. A minimum goal for a logical encoded qubit memory is to show a lower error rate than the unencoded qubits used to encode it. For a bare idle error rate of $q=10^{-4}$ the penalty for adding electronics constraints is that the gate probability of failure, p, must be approximately five times better than the unconstrained case to show benefit. In addition the maximum benefit that can be achieved as the gate error rate, p, approaches zero for any value of q is approximately 3 times worse when electronics constraints are considered.

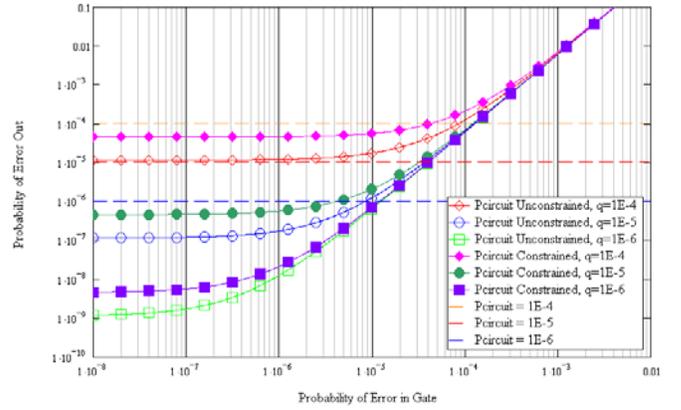

Fig. 7. Pessimistic bound of circuit failure probability as a function of individual gate error probability. A comparison is made for a schedule that includes electronics constraints (i.e., increased idles) and is free of local constraints for fixed idle error rates of $10^{-4}$ (triangles), $10^{-5}$ (circles), and $10^{-6}$ (squares). These error probabilities are the same order of magnitude as would be expected for 30 ns idle steps, defined by the classical electronics design, and measured T2 times of electron spins in the bulk, 60 ms [7], and near an oxide surface, ~0.3 $\mu$s [9]. Horizontal lines indicate at what values of p the circuit performs better than the bare idle error rate q.

In general, the dependence of the quantum circuit error probability on electronics manifests itself in this calculation through 1) the increase in idles, M; 2) the probability of error during an idle, q, which is a function of the quantum clock $T_{qclk}$ also defined by the electronics; and 3) the choice of QEC code, which is influenced by the choice of physical qubit, its native gate set and it's available lay-out defined by the electronics. The specific choice of physical qubit profoundly influences these parameters through available lay-out, reduced parallelism, forced operation within a cryostat having limited I/O lines, and limits on speed of gates due to the physical qubits requirements for electronics accuracy. We conclude that two important figures of merit that can be identified from the micro-architecture analysis are, the quantum clock period and the increase in the number of quantum clock idles due to parallelism penalties that result from the definition of constraints produced by the electronics.

## VI. SUMMARY

We discuss the impact of classical electronics on many aspects of a micro-architecture, a quantum error corrected memory using silicon double quantum dot qubits. The effects on the quantum error correction circuit discussed include: 1) reducing the number of parallel quantum operations possible; 2) setting the quantum clock time; 3) dictating the lay-out and possible code choice at the qubit chip level; and 4) electronics limits on the minimum error rate of certain critical quantum gates. Timing and voltage accuracy limits in the electronics leads to a limit on the shortest gate time. The shortest gate time has implications for the error probability related to idle qubits because the scheduling of gate operations is more conveniently parsed in terms of time intervals, ticks, based on the shortest gate time. For example, it is potentially more relevant to use the ratio $T_{Qclk} / T_2$ to estimate whether a system will approach the pseudo-threshold for a particular gate in a quantum error correction circuit rather than the more common figure of merit used in the literature, $T_{gate} / T_2$ [14]. $T_{Qclk}$ is a system parameter defined by the micro-architecture and electronics constraints. Furthermore, the reduction in parallel gate operations manifests itself as a penalty in additional idle ticks, M, in the QEC schedule leading to a factor of approximately 2 increase in idle steps relative to a constraint free schedule. The increase in probability of a circuit error due to this increase in idle steps depends strongly on the idle error probability, but can become a significant shift for experimentally relevant decoherence times and the system determined system clock period. We find that the details of this QEC micro-architecture can have a very strong influence on the overall performance of the quantum circuitry, although it is not typically captured in larger scale architecture analysis. We also add that this analysis, though specific to a particular solid-state implementation of a BS9(21) architecture comprised of Si-DQD's, is relevant and can be applied to other architectures [11], [38], [39], solid state qubit systems (such as GaAs and SiGe [4], [40], [41]), and qubit implementations [39], [42].

## VII. ACKNOWLEDGMENTS

The authors are especially grateful for discussions with Dr. Wayne M. Witzel, Dr Uzoma Onunkwo, and Jason R Hamlet of Sandia National Laboratories for useful discussions regarding quantum noise modeling, error probability calculations, and cryogenic cooling.